\documentclass[aps,prd,twocolumn,nofootinbib,showpacs]{revtex4}

\usepackage{graphicx,color}
\usepackage[colorlinks=true, linkcolor=blue, citecolor=blue, urlcolor=blue]{hyperref}

\begin{document}

\title{QCD improved top-quark decay at next-to-next-to-leading order}

\author{Rui-Qing Meng$^1$}

\author{Sheng-Quan Wang$^1$}
\email[email:]{sqwang@cqu.edu.cn}

\author{Ting Sun$^1$}

\author{Chao-Qin Luo$^1$}

\author{Jian-Ming Shen$^{2}$}

\author{Xing-Gang Wu$^3$}
\email[email:]{wuxg@cqu.edu.cn}

\address{$^1$Department of Physics, Guizhou Minzu University, Guiyang 550025, P.R. China}
\address{$^2$School of Physics and Electronics, Hunan University, Changsha 410082, P.R. China}
\address{$^3$Department of Physics, Chongqing Key Laboratory for Strongly Coupled Physics, Chongqing University, Chongqing 401331, P.R. China}

\date{\today}

\begin{abstract}

We analyse the top-quark decay at the next-to-next-to-leading order (NNLO) in QCD by using the Principle of Maximum Conformality (PMC) which provides a systematic way to eliminate renormalization scheme and scale ambiguities in perturbative QCD predictions. The PMC renormalization scales of the coupling constant $\alpha_s$ are determined by absorbing the non-conformal $\beta$ terms that govern the behavior of the running coupling by using the Renormalization Group Equation (RGE). We obtain the PMC scale $Q_\star=15.5$ GeV for the top-quark decay, which is an order of magnitude smaller than the conventional choice $\mu_r=m_t$, reflecting the small virtuality of the QCD dynamics of the top-quark decay process. Moreover, due to the non-conformal $\beta$ terms disappear in the pQCD series, there is no renormalon divergence and the NLO QCD correction term is greatly increased while the NNLO QCD correction term is suppressed compared to the conventional results obtained at $\mu_r=m_t$. By further including the next-to-leading (NLO) electroweak corrections, the finite $W$ boson width and the finite bottom quark mass, we obtain the top-quark total decay width $\Gamma^{\rm tot}_t=1.3112^{+0.0190}_{-0.0189}$ GeV, where the error is the squared averages of the top-quark mass $\Delta m_t=\pm0.7$ GeV, the coupling constant $\Delta \alpha_s(M_Z)=\pm0.0009$ and the estimation of unknown higher-order terms using the PAA method with [N/M]=[1/1]. The PMC improved predictions for the top-quark decay are complementary to the previous PMC calculations for top-quark pair production and helpful for detailed studies of properties of the top-quark.

\end{abstract}

\maketitle

\section{Introduction}
\label{sec:1}

The top-quark was discovered in 1995 by the CDF and D0 Collaborations~\cite{CDF:1995wbb,D0:1995jca} and it is the heaviest fermion of the Standard Model (SM). Due to the large mass of the top-quark, it is the only particle that decays without hadronization in elementary particles. Detailed studies of properties of the top-quark such as its mass, its production and the structure of its couplings to other elementary particles plays a crucial role for understanding the nature of electroweak symmetry breaking and searching for new physics beyond the SM. Many of such studies require a good understanding of the decay of the top-quark, thus a detailed study of its decay process is highly desirable.

Significant efforts have been made on theoretical calculations for the top-quark decay process. The next-to-leading order (NLO) QCD corrections to the top-quark decay have been computed in Refs.\cite{Jezabek:1988iv,Czarnecki:1990kv,Li:1990qf,Jezabek:1993wk}. The NLO QCD corrections to the top-quark production and decay are incorporated consistently, which has been extensively studied~\cite{Bigi:1986jk,Campbell:2004ch,Campbell:2005bb,Campbell:2012uf,Melnikov:2009dn,Bernreuther:2010ny,Melnikov:2011ta,Melnikov:2011qx}. The next-to-next-to-leading order (NNLO) QCD corrections to the total width of the top-quark were calculated in Refs.\cite{Czarnecki:1998qc,Chetyrkin:1999ju,Blokland:2004ye,Blokland:2005vq}. In recent years, fully differential calculations to the top-quark decay rate at NNLO were performed in Refs.\cite{Gao:2012ja,Brucherseifer:2013iv}, as well as the calculation of the combination of the top-quark production and decay beyond NLO were given in Refs.\cite{Berger:2016oht,Gao:2017goi,Berger:2017zof,Liu:2018gxa,Czakon:2020qbd}. Experimentally, the Tevatron and LHC experiments have measured the total width of the top-quark decay using different methods. For example, the first direct measurement comes from the ATLAS Collaboration at $\sqrt{s}=8$ TeV, yielding $\Gamma_t=1.76\pm0.33^{+0.79}_{-0.68}$ GeV~\cite{ATLAS:2017vgz}. A novel method for directly measuring the top-quark decay using events away from the resonance peak, gives $\Gamma_t=1.28\pm0.30$ GeV~\cite{Herwig:2019obz}. The most precise value $\Gamma_t=1.36\pm0.02^{+0.14}_{-0.11}$ GeV~\cite{CMS:2014mxl} was presented by the CMS Collaboration with the assumption of the branching fraction $\mathcal{BR}(t\rightarrow Wq)=1$. The Particle Data Group (PDG) was reported the world average: $\Gamma_t=1.42^{+0.19}_{-0.15}$ GeV~\cite{PDG:2020}.

Extensively studying the properties of the top-quark calls for a precise theoretical prediction for the width of the top-quark decay. According to the conventional practice, there are renormalization scheme-and-scale ambiguities for fixed-order pQCD predictions. The choice of the renormalization scale is arbitrary, and one usually choose the renormalization scale as the top-quark mass $\mu_r=m_t$ to eliminate the large logarithmic terms $\ln(m^2_t/\mu^2_r)$ of pQCD series and then estimates the uncertainty by varying the renormalization scale over an arbitrary range, e.g., $\mu_r\in[m_t/2, 2m_t]$ for the prediction of the top-quark decay. As a matter of fact the variation of the renormalization scale is only sensitive to the non-conformal $\beta$ terms but not to the conformal terms and moreover we do not know how wide a range the renormalization scale should vary in order to achieve reliable predictions for its uncertainty. The renormalization scale ambiguity becomes one of the most important systematic errors for pQCD predictions.

The Principle of Maximum Conformality (PMC)~\cite{Brodsky:2011ta, Brodsky:2012rj, Brodsky:2011ig, Mojaza:2012mf, Brodsky:2013vpa} has been proposed to eliminate the renormalization scheme-and-scale ambiguities in pQCD predictions. The PMC provides underlying principle for the Brodsky-Lepage-Mackenzie (BLM) method and extends the BLM method~\cite{Brodsky:1982gc} to all orders. We have shown that a comprehensive, self-consistent pQCD explanation for the top-quark pair hadroproduction can be obtained~\cite{Brodsky:2012rj, Brodsky:2012sz, Brodsky:2012ik, Wang:2014sua, Wang:2015lna}, and a precise determination of the top-quark mass can be achieved~\cite{Wang:2017kyd, Wang:2020mel} by applying the PMC method. The PMC scales are determined by absorbing the $\beta$ terms that govern the behavior of the running $\alpha_s$ via the Renormalization Group Equation (RGE) and thus we can determine the effective coupling constant for the top-quark pair hadroproduction. The PMC prediction satisfies the requirement of the renormalization group invariance~\cite{Brodsky:2012ms, Wu:2014iba, Wu:2019mky}. Since the $\beta$ terms do not appear in the perturbative QCD series, there is no renormalon divergence by using the PMC. The predicted top-quark pair production cross sections agree with precise experimental data, and the large discrepancies of the top-quark forward-backward asymmetries between SM estimations and experimental data are greatly reduced~\cite{Brodsky:2012rj, Brodsky:2012sz, Brodsky:2012ik, Wang:2014sua, Wang:2015lna}. After having higher-order QCD corrections, the predictions obtained by using conventional scale setting are also in agreement with experimental data~\cite{Czakon:2013goa, Czakon:2015owf, Czakon:2017lgo}. A detailed PMC analysis for the top-quark decay process is desirable, since it is an important ingredient in the calculation of the combination of the top-quark pair production and decay.

Our goal in this paper is to provide a detailed PMC analysis for the top-quark decay. The remaining sections of this paper are organized as follows. In Sec.~\ref{sec:2}, we present our calculation technology for applying the PMC to the top-quark decay at NNLO. We give numerical results and discussions for the top-quark decay in Sec.~\ref{sec:3}. The paper is concluded with a summary in Sec.~\ref{sec:4}

\section{PMC scale setting for the top-quark decay}
\label{sec:2}

The top-quark decay precess is almost completely dominated by the $t\rightarrow bW$, with the subsequent decays of the $W$ bosons into charged leptons, or into quarks. QCD corrections to this precess at NNLO can be written as
\begin{eqnarray}
\Gamma_t=\Gamma^{\rm LO}_t\left[1+c_1\,a_s(\mu_r)+c_2(\mu_r)\,a^2_s(\mu_r)+{\cal O}(\alpha^3_s)\right],
\label{Gamma_NNLO}
\end{eqnarray}
where $a_s(\mu_r)=\alpha_s(\mu_r)/(4\pi)$, and $\mu_r$ stand for the renormalization scale. The decay width at leading order (LO) is given by
\begin{eqnarray}
\Gamma^{\rm LO}_t=\frac{G_F\,|V_{tb}|^2\,m_t^3}{8\,\pi\,\sqrt{2}}\left(1-3\,w^2+2\,w^3\right),
\end{eqnarray}
where $w={m^2_W}/{m^2_t}$, $G_F$ is the Fermi constant, $|V_{tb}|$ denotes Cabibbo-Kobayashi-Maskawa (CKM) matrix element, and $m_t$ is the mass of the top-quark.

The coefficients $c_1$ and $c_2(\mu_r)$ are for the NLO and NNLO QCD corrections, respectively, which have been extensively studied in the literature. The NNLO coefficient $c_2(\mu_r)$ can be divided into $n_f$-dependent and $n_f$-independent parts,
\begin{eqnarray}
c_2(\mu_r)=c_{2,0}(\mu_r)+c_{2,1}(\mu_r)\,n_f,
\end{eqnarray}
where $n_f$ is the number of active quark flavours, which is related to the $\beta_0$ term by $\beta_0=11-2/3\,n_f$. The perturbative coefficients can be further divided into conformal terms and non-conformal terms~\cite{Mojaza:2012mf, Brodsky:2013vpa} and the top-quark decay width $\Gamma_t$ in Eq.(\ref{Gamma_NNLO}) changes to
\begin{eqnarray}
\Gamma_t&=&\Gamma^{\rm LO}_t\left[1+r_{1,0}\,a_s(\mu_r)+\left(r_{2,0}(\mu_r)\right.\right. \nonumber\\
&&\left.\left.+r_{2,1}(\mu_r)\,\beta_0\right)\,a^2_s(\mu_r)+{\cal O}(\alpha^3_s)\right].
\label{Gamma_NNLOb}
\end{eqnarray}
For our present NNLO analysis, the coefficients
\begin{eqnarray}
r_{1,0}&=&c_1, \nonumber\\
r_{2,0}(\mu_r)&=&c_{2,0}(\mu_r)+\frac{33}{2}\,c_{2,1}(\mu_r)
\end{eqnarray}
stand for the conformal terms, and the coefficient
\begin{eqnarray}
r_{2,1}(\mu_r)=-\frac{3}{2}\,c_{2,1}(\mu_r)
\end{eqnarray}
is the non-conformal term.

We shall adopt the PMC single-scale method~\cite{Shen:2017pdu} for our analysis, and an overall scale can be determined for the top-quark decay process. The NNLO coefficient is adopted from Refs.\cite{Blokland:2004ye,Blokland:2005vq}, which is confirmed by the fully differential calculations~\cite{Gao:2012ja,Brucherseifer:2013iv}. After applying PMC scale setting to the top-quark decay width in Eq.(\ref{Gamma_NNLOb}), we obtain
\begin{eqnarray}
\Gamma_t&=&\Gamma^{\rm LO}_t\left[1+r_{1,0}\,a_s(Q_\star)+r_{2,0}(\mu_r)\,a^2_s(Q_\star)+{\cal O}(\alpha^3_s)\right].
\label{Gamma_NNLOpmc}
\end{eqnarray}
Here, $Q_\star$ is the PMC scale, which is determined by requiring all non-conformal terms vanish. At present NNLO level, the PMC scale is given by
\begin{eqnarray}
Q_\star=\mu_r\,\exp\left[-\frac{r_{2,1}(\mu_r)}{2\,r_{1,0}}+{\cal O}(\alpha_s)\right].
\end{eqnarray}

After applying the PMC, only the conformal terms remain in pQCD series. Thus, the resulting perturbative series matches the conformal series of the conformal theory. At present NNLO level, the PMC scale $Q_\star$ and the conformal coefficient $r_{2,0}$ only formally depend on the choice of the renormalization scale $\mu_r$, and their values are independent of the choice of renormalization scale $\mu_r$. The renormalization scale dependence is eliminated for the top-quark decay width in Eq.(\ref{Gamma_NNLOpmc}).

The previous NNLO QCD calculation~\cite{Blokland:2004ye,Blokland:2005vq,Gao:2012ja,Brucherseifer:2013iv} are given by using the top-quark pole mass. In this paper, we also give a PMC analysis for the top-quark decay process in the pole mass scheme. It is noted that by using the $\overline{\rm MS}$ mass of the top-quark, the convergence of the pQCD series is greatly improved for both the top-quark pair hadroproduction (e.g,~\cite{Catani:2020tko}) and the top-quark decay (e.g,~\cite{Beneke:1994qe}). It is of interest to analyze these processes in the $\overline{\rm MS}$ mass scheme by using the PMC method. However, in the $\overline{\rm MS}$ mass scheme, the scale evolution of the running coupling $\alpha_s(\mu_r)$ and the $\overline{\rm MS}$ mass $m(\mu_r)$ are governed by a general renormalization group equation involving both the $\beta$ function and the quark mass anomalous dimension $\gamma_m$, and the $n_f$-terms from both the QCD $\beta$-function and the anomalous dimension are entangled with each other. To give a PMC analysis in the $\overline{\rm MS}$ mass scheme, one needs to distinguish the source of the $n_f$-terms correctly. For example, a detailed analysis in the $\overline{\rm MS}$ mass scheme is given in Ref.\cite{Huang:2022rij}.

\section{Numerical results and discussions}
\label{sec:3}

To do numerical calculations, the two-loop $\overline{\rm MS}$ scheme QCD coupling is evaluated from $\alpha_s(M_Z)=0.1179$, the top-quark pole mass $m_t=172.5$ GeV~\cite{PDG:2020}, the $W$ boson mass $m_W=80.385$ GeV, the Fermi constant $G_F=1.16638\times10^{-5}$ GeV$^{-2}$ and the CKM matrix element $|V_{tb}|=1$~\cite{Gao:2012ja}.

\begin{table} [htb]
\begin{tabular}{|c||c|c|c|c|c|}
\hline
~~ ~~  & ~$\mu_r$~ & ~$\Gamma^{\rm LO}_t$~ & ~$\delta\Gamma^{\rm NLO}_t$~ & ~$\delta\Gamma^{\rm NNLO}_t$~ & ~$\Gamma^{\rm NNLO}_t$~ \\
\hline
~~ ~~  & ~$m_t/2$~ & ~1.4806~ & ~-0.1394~ & ~-0.0234~ & ~1.3179~ \\
~Conv.~& ~$m_t$~   & ~1.4806~ & ~-0.1261~ & ~-0.0306~ & ~1.3239~  \\
~~ ~~  & ~$2m_t$~  & ~1.4806~ & ~-0.1161~ & ~-0.0357~ & ~1.3288~  \\
\hline
~PMC~ & ~~         & ~1.4806~ & ~-0.1892~ & ~0.0207~  & ~1.3122~ \\
\hline
\end{tabular}
\caption{The LO decay width $\Gamma^{\rm LO}_t$ together with the NLO and NNLO QCD correction terms $\delta\Gamma^{\rm NLO}_t$ and $\delta\Gamma^{\rm NNLO}_t$ for the top-quark decay using the conventional (Conv.) and PMC scale settings. Decay width are shown in unit of GeV.
\label{tab1} }
\end{table}

The top-quark decay width $\Gamma^{\rm LO}_t$ at LO is free from QCD strong interaction and provides dominant contributions. The QCD interaction occurs starting from NLO, and the corrections are proportional to the QCD coupling $\alpha_s$. According to the conventional scale-setting method, the renormalization scale of $\alpha_s$ is usually set to the top-quark mass $\mu_r=m_t$, and its uncertainty is estimated by varying the scale over an arbitrary range; e.g., $\mu_r\in[m_t/2, 2m_t]$. In Table \ref{tab1} we present the LO decay width $\Gamma^{\rm LO}_t$ together with the NLO and NNLO QCD correction terms $\delta\Gamma^{\rm NLO}_t$ and $\delta\Gamma^{\rm NNLO}_t$ for the top-quark decay using the conventional (Conv.) and PMC scale settings. At present NNLO level, the scale uncertainty for the total QCD correction term $\delta\Gamma^{\rm NLO}_t+\delta\Gamma^{\rm NNLO}_t$ is $\sim[-3.8\%,+3.1\%]$ for $\mu_r\in[m_t/2,2m_t]$. Due to the LO decay width $\Gamma^{\rm LO}_t$ provides dominant contributions, the scale uncertainty for the top-quark decay width $\Gamma^{\rm NNLO}_t$ is suppressed to $\sim[-0.5\%,+0.4\%]$. The $\delta\Gamma^{\rm NLO}_t$ increases with the increasing of the scale $\mu_r$; the $\delta\Gamma^{\rm NNLO}_t$ decreases with the increasing of the scale $\mu_r$. Thus, the scale uncertainties of $\delta\Gamma^{\rm NLO}_t$ and $\delta\Gamma^{\rm NNLO}_t$ cancel each other out, which leads to a small scale uncertainty for $\delta\Gamma^{\rm NLO}_t+\delta\Gamma^{\rm NNLO}_t$. Compared to the total QCD correction $\delta\Gamma^{\rm NLO}_t+\delta\Gamma^{\rm NNLO}_t$, the scale uncertainty is rather large for each QCD correction term, i.e., the scale uncertainties are $\sim[-10.5\%,+7.9\%]$ for the $\delta\Gamma^{\rm NLO}_t$ and $\sim[+23.5\%, -16.7\%]$ for the $\delta\Gamma^{\rm NNLO}_t$ for $\mu_r\in[m_t/2,2m_t]$.

In the case of conventional scale setting, one cannot decide what is the exact QCD correction terms for each perturbative order, and the renormalization scale uncertainty becomes one of the most important systematic errors. After applying PMC scale setting, the QCD correction terms are almost fixed to be $-0.1892$ GeV for $\delta\Gamma^{\rm NLO}_t$ and $0.0207$ GeV for $\delta\Gamma^{\rm NNLO}_t$ for any choice of the renormalization scale $\mu_r$. Thus, the renormalization scale uncertainty of conventional scale setting is eliminated.

It is noted that by fixing the scale $\mu_r=m_t$, the relative importance of the NLO and NNLO QCD correction terms
\begin{eqnarray}
\delta\Gamma^{\rm NLO}_t/\Gamma^{\rm LO}_t&\sim&-8.6\%, \nonumber\\
\delta\Gamma^{\rm NNLO}_t/\Gamma^{\rm LO}_t&\sim&-2.1\%
\end{eqnarray}
are given in Ref.\cite{Gao:2012ja}. Using the same input parameters, our conventional results agree with those of Ref.\cite{Gao:2012ja}. However, Table \ref{tab1} shows that the NLO and NNLO QCD correction terms are changed to
\begin{eqnarray}
\delta\Gamma^{\rm NLO}_t/\Gamma^{\rm LO}_t&\sim&-9.4\%, \nonumber\\
\delta\Gamma^{\rm NNLO}_t/\Gamma^{\rm LO}_t&\sim&-1.6\%
\end{eqnarray}
for $\mu_r=m_t/2$, which implies that the convergence of the pQCD series is improved. The NLO and NNLO QCD correction terms are changed to
\begin{eqnarray}
\delta\Gamma^{\rm NLO}_t/\Gamma^{\rm LO}_t&\sim&-7.8\%, \nonumber\\
\delta\Gamma^{\rm NNLO}_t/\Gamma^{\rm LO}_t&\sim&-2.4\%
\end{eqnarray}
for $\mu_r=2m_t$, which implies a slower convergence of the pQCD series. Thus, in the case of conventional scale setting, one cannot decide the intrinsic convergence of the pQCD series, and a poor convergence of the pQCD series may be caused by the improper choice of the renormalization scale.

After applying PMC scale setting, Table \ref{tab1} shows that the PMC results for the NLO and NNLO QCD correction terms are fixed to
\begin{eqnarray}
\delta\Gamma^{\rm NLO}_t/\Gamma^{\rm LO}_t&\sim&-12.8\%, \nonumber\\
\delta\Gamma^{\rm NNLO}_t/\Gamma^{\rm LO}_t&\sim&1.4\%
\end{eqnarray}
for any choice of the scale $\mu_r$. Due to the absorption of the non-conformal terms, the NLO QCD correction term is greatly increased while the NNLO QCD correction term is suppressed compared to the conventional results obtained in the range of $\mu_r\in[m_t/2,2m_t]$.

It is noted that the scale-independent NNLO conformal coefficient $r_{2,0}(\mu_r)$ is significantly different from the conventional NNLO coefficient $c_2(\mu_r)$, i.e., $r_{2,0}(\mu_r)=85.0$, and $c_2(\mu_r)=-(282.2\pm105.7)$ for $\mu_r\in[m_t/2,2m_t]$. Thus, the NNLO QCD correction term provides a negative value using conventional scale setting; it becomes a positive value after using the PMC. By fixing the scale $\mu_r=m_t$, the conventional prediction for the top-quark decay width is $\Gamma^{\rm NNLO}_t|_{\rm Conv.}=1.3239^{+0.0049}_{-0.0060}$ GeV, where the error is caused by $\mu_r\in[m_t/2,2m_t]$. The scale-independent PMC prediction for the top-quark decay width is $\Gamma^{\rm NNLO}_t|_{\rm PMC}=1.3122$ GeV, which is smaller than the conventional predictions obtained in $\mu_r\in[m_t/2,2m_t]$.

\begin{figure} [htb]
\centering
\includegraphics[width=0.40\textwidth]{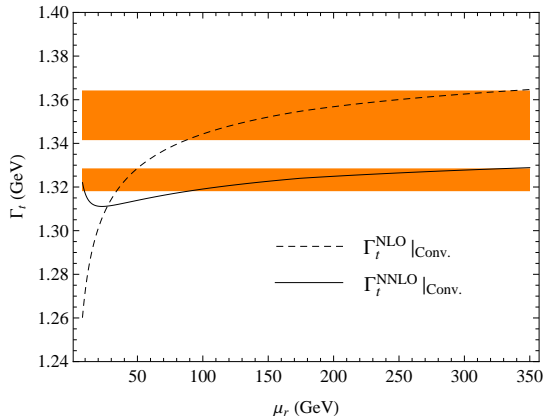}
\caption{The top-quark decay widths $\Gamma_t$ versus the renormalization scale $\mu_r$ using conventional (Conv.) scale setting, where the dashed and solid lines stand for the conventional predictions $\Gamma^{\rm NLO}_t|_{\rm Conv.}$ at NLO and $\Gamma^{\rm NNLO}_t|_{\rm Conv.}$ at NNLO, respectively. Two orange bands stand for the estimation of the unknown higher-order QCD terms by using the conventional method of $\mu_r\in[m_t/2,2m_t]$. }
\label{topdecayConv}
\end{figure}

\begin{figure} [htb]
\centering
\includegraphics[width=0.40\textwidth]{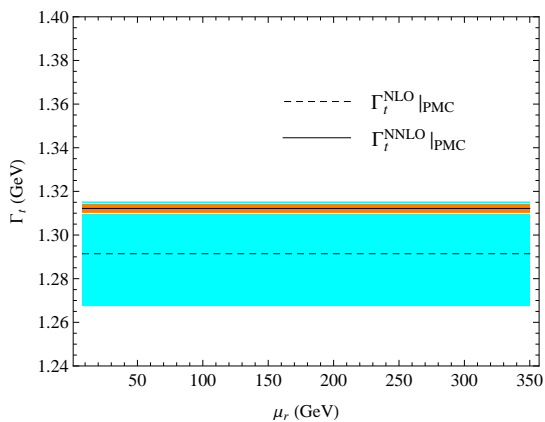}
\caption{The top-quark decay widths $\Gamma_t$ versus the renormalization scale $\mu_r$ using PMC scale setting, where the dashed and solid lines stand for the PMC predictions $\Gamma^{\rm NLO}_t|_{\rm PMC}$ at NLO and $\Gamma^{\rm NNLO}_t|_{\rm PMC}$ at NNLO, respectively. Two error bands stand for the estimation of the unknown higher-order QCD terms by using the Pad$\acute{e}$ approximation approach. }
\label{topdecayPMC}
\end{figure}

More explicitly, in Figs.(\ref{topdecayConv}) and (\ref{topdecayPMC}) we present the top-quark decay widths $\Gamma_t$ at NLO and NNLO versus the renormalization scale $\mu_r$ using the conventional and PMC scale settings. In the case of conventional scale setting, we can see from figure (\ref{topdecayConv}) that the top-quark decay width $\Gamma_t$ at the NLO QCD correction depends heavily on the choice of the renormalization scale $\mu_r$, whereas the scale dependence becomes weaker by the inclusion of the NNLO QCD correction. This observation is consistent with the conventional expectation that the scale dependence is progressively decreased by the inclusion of higher-order QCD calculations. Figure (\ref{topdecayConv}) also shows clearly that by varying the scale $\mu_r\in[m_t/2,2m_t]$, the NNLO calculation does not overlap with the NLO prediction. Thus the estimate of unknown higher-order QCD terms by $\mu_r\in[m_t/2,2m_t]$ is unreliable for the top-quark decay process. In fact, simply varying the scale is only sensitive to the $\beta$ terms, but not to the conformal terms. Some typical examples such as the hadroproduction of the Higgs boson at NLO~\cite{Anastasiou:2002yz, Ravindran:2003um, Wang:2016wgw}, the event shape observables in electron-positron annihilation~\cite{GehrmannDeRidder:2007hr, Weinzierl:2009ms} illustrate the unreliability of error estimates using this conventional method.

After using the PMC, as show by figure (\ref{topdecayPMC}), the top-quark decay widths at NLO and NNLO are almost flat versus the renormalization scale $\mu_r$; the scale dependence of both the QCD correction terms at each perturbative order and the total QCD correction are simultaneously eliminated. Simply varying the scale to estimate the unknown higher-order QCD terms is not applicable to the PMC predictions, since the PMC scales are determined unambiguously by the non-conformal terms, and the variation of the PMC scales would break the renormalization group invariance and then lead to an unreliable prediction~\cite{Wu:2014iba}.

We adopt the Pad$\acute{e}$ approximation approach (PAA)~\cite{Basdevant:1972fe, Samuel:1992qg, Samuel:1995jc} for the estimate of unknown higher-order QCD terms. The PAA offers a feasible conjecture that yields the unknown $(n+1)_{\rm th}$-order terms from the given $n_{\rm th}$-order perturbative series, and a $[N/M]$-type approximant $\rho_n^{[N/M]}$ for $\rho_n=\sum_{i=0}^{n(\geq 1)}C_i x^i$ is defined as
\begin{eqnarray}
\rho^{[N/M]}_n &=& \frac{b_0+b_1 x + \cdots + b_N x^N} {1 + c_1 x + \cdots + c_M x^M} \nonumber \\
               &=& \sum_{i=0}^{n} C_i x^i + C_{n+1} \; x^{n+1} +\cdots,
\end{eqnarray}
where the parameter $M\geq 1$ and $N+M=n$. The known coefficients $C_{i(\leq n)}$ determine the parameters $b_{i\in[0,N]}$ and $c_{j\in[1,M]}$, which inversely predicts a reasonable value for the next uncalculated coefficient $C_{n+1}$~\cite{Du:2018dma}. Applying the Pad$\acute{e}$ approach to the Eq.(\ref{Gamma_NNLOpmc}), the estimated NNLO coefficient is $r_{2,0}^{\rm PAA}=r^2_{1,0}$ for $[N/M]=[0/1]$; the estimated NNNLO coefficient is $r_{3,0}^{\rm PAA}=r^2_{2,0}/r_{1,0}$ for $[N/M]=[1/1]$. Thus, the uncertainty from unknown higher-order terms could be estimated by $\Delta\Gamma^{\rm NLO}_t=\pm r_{2,0}^{\rm PAA}\,a^{2}_s(Q_\star)$ for the NLO PMC result, and $\Delta\Gamma^{\rm NNLO}_t=\pm r_{3,0}^{\rm PAA}\,a^{3}_s(Q_\star)$ for the NNLO PMC result. Finally, we obtain $\Delta\Gamma^{\rm NLO}_t=\pm0.0242$ GeV and $\Delta\Gamma^{\rm NNLO}_t=\pm725.4077\,a^{3}_s(Q_\star)=\pm0.0023$ GeV~\footnote{The estimated NNNLO coefficient is $r_{3,0}^{\rm PAA}=-r^3_{1,0}+2\,r_{1,0}\,r_{2,0}$ for $[N/M]=[0/2]$, which leads to $\Delta\Gamma^{\rm NNLO}_t=\pm705.6404\,a^{3}_s(Q_\star)=\pm0.0022$ GeV, and is very close to the case of $[N/M]=[1/1]$.}. The NNLO PMC results are well within the error bars predicted from the NLO PMC calculations for the top-quark decay based on PMC+PAA calculation, as shown clearly by figure (\ref{topdecayPMC}).

At present NNLO level, we obtain one scale for the top-quark decay and the determined PMC scale is
\begin{eqnarray}
Q_\star=15.5 ~\rm {GeV}.
\end{eqnarray}
The PMC scale is independent of the renormalization scale $\mu_r$, and it is an order of magnitude smaller than the conventional choice $\mu_r=m_t$, reflecting the small virtuality of the QCD dynamics of the top-quark decay process. In addition, as shown by figure (\ref{topdecayConv}), the top-quark decay width at NNLO first decreases and then increases with increasing scale $\mu_r$ using conventional scale setting, achieving its minimum value at $\mu_r\sim 23$ GeV. If one chooses to replace the conventional choice $\mu_r=m_t$ with the small scale $\mu_r\ll m_t$ (especially around $23$ GeV), the pQCD convergence of the top-quark decay will be greatly improved, as well as the resulting conventional prediction decreases and close to the scale-independent PMC prediction. Thus, the effective momentum flow for the top-quark decay process should be $\mu_r\ll m_t$, far lower than the conventionally suggested $\mu_r=m_t$. Some other examples such as the bottom quark forward-backward asymmetry~\cite{Wang:2020ell} and the event shape observables~\cite{Wang:2019isi} in electron-positron annihilations also show that the effective renormalization scale is very small compared to the scale of conventional choice.

\begin{table} [htb]
\begin{tabular}{|c|c|c|c|c|c|}
\hline
 $m_t$        & $\Gamma^{\rm NNLO}_t|_{\rm PMC}$  & $\delta_f^b$ & $\delta_f^W$  & $\delta^{\rm NLO}_{\rm EW}$ & Total \\
\hline
~172.5~       & ~$1.3122$~ & ~-0.0038~ & ~-0.0221~  & ~0.0249~ & ~1.3112~   \\
\hline
~173.5~       & ~$1.3392$~ & ~-0.0039~ & ~-0.0225~  & ~0.0255~ & ~1.3383~   \\
\hline
\end{tabular}
\caption{The PMC results of the top-quark decay widths $\Gamma^{\rm NNLO}_t|_{\rm PMC}$ together with the corrections from the finite bottom quark mass $\delta_f^b$, the finite $W$ boson width $\delta_f^W$ and the NLO electroweak corrections $\delta^{\rm NLO}_{\rm EW}$ (in unit GeV) for $m_t=172.5$ and $173.5$ GeV. These corrections $\delta_f^b$, $\delta_f^W$ and $\delta^{\rm NLO}_{\rm EW}$ are taken from Ref.\cite{Gao:2012ja}.
\label{tab3} }
\end{table}

In order to provide a reliable prediction for the top-quark decay, we need to take into account other corrections such as the effect of finite bottom quark mass and finite $W$ boson width, as well as electroweak corrections. In Table \ref{tab3} we present the PMC results of the top-quark decay widths together with the corrections from the finite bottom quark mass, the finite $W$ boson width and the NLO electroweak corrections for $m_t=172.5$ and $173.5$ GeV. These corrections are taken from Ref.\cite{Gao:2012ja}. Since the corrections from the finite bottom quark mass and the finite $W$ boson width provide negative values while the NLO electroweak correction provides a positive value, their contributions cancel out greatly to the top-quark decay.

\begin{figure} [htb]
\centering
\includegraphics[width=0.40\textwidth]{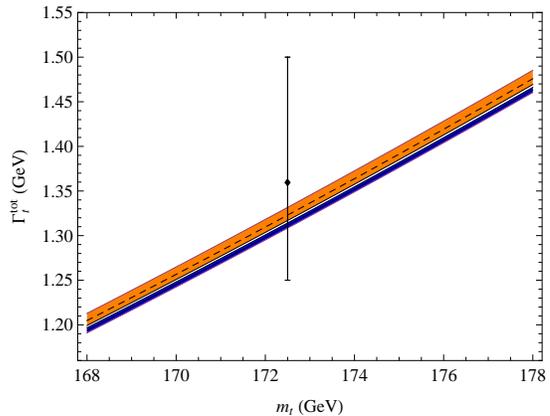}
\caption{The top-quark total decay width $\Gamma^{\rm tot}_t$ versus the top-quark mass $m_t$, where the solid line and the blue band represent the PMC predictions and the dashed line and the orange band stand for the conventional predictions. As a comparison, the most precise experimental measurements~\cite{CMS:2014mxl} is also presented. }
\label{topdecaymt}
\end{figure}

Finally, we obtain reliable predictions for the top-quark total decay width
\begin{eqnarray}
\Gamma^{\rm tot}_t&=&1.3112\pm 0.0016\pm 0.0023 ~\rm {GeV}
\end{eqnarray}
for $m_t=172.5$ GeV, and
\begin{eqnarray}
\Gamma^{\rm tot}_t&=&1.3383^{+0.0016}_{-0.0017}\pm 0.0023 ~\rm {GeV}
\end{eqnarray}
for $m_t=173.5$ GeV. Here, the first error comes from the coupling constant $\Delta \alpha_s(M_Z)=\pm0.0009$~\cite{PDG:2020} and the second error is caused by the estimation of unknown higher-order terms using the PAA method. In the case of conventional scale setting, by including the corrections from the finite bottom quark mass $\delta_f^b$, the finite $W$ boson width $\delta_f^W$ and the NLO electroweak corrections $\delta^{\rm NLO}_{\rm EW}$, the top-quark total decay width is $\Gamma^{\rm tot}_t=1.3229$ GeV for $\mu_r=m_t=172.5$ GeV. It is noted that the top-quark mass uncertainties are larger than the discrepancies between conventional and PMC scale settings to the perturbative expansion. The top-quark total decay width depends heavily on the top-quark mass, and thus the theoretical error is dominated by the $m_t$. More explicitly, we show the top-quark total decay width $\Gamma^{\rm tot}_t$ versus the top-quark mass $m_t$ in Fig.(\ref{topdecaymt}). The most precise experimental measurements (about $10\%$ uncertainty)~\cite{CMS:2014mxl} is also presented as a comparison. Currently, the experimental measurements have relatively large uncertainties. The experimental measurements and the theoretical predictions of the PMC and conventional scale settings are in agreement.

\section{Summary }
\label{sec:4}

Using conventional scale setting, the renormalization scale is usually set to the top-quark mass $m_t$, and its uncertainty is estimated by varying the scale over an arbitrary range for the pQCD prediction of the top-quark decay. Such conventional scale-setting procedure introduces an inherent scheme-and-scale dependence. One cannot decide what are the exact QCD correction terms for each perturbative order as well as the intrinsic convergence of the pQCD series. A improper choice of the renormalization scale can lead to a poor convergence of the pQCD series. The renormalization scale uncertainty becomes one of the most important errors for pQCD predictions.

After applying PMC scale setting, the PMC renormalization scales are determined unambiguously by absorbing all of the non-conformal terms, and only the conformal terms remain in pQCD series. The renormalization scale uncertainty is eliminated for pQCD predictions. The intrinsic convergence of the pQCD series of the top-quark decay is greatly improved due to the renormalon divergences are eliminated. The determined PMC scale for the top-quark decay is $Q_\star=15.5$ GeV, which is an order of magnitude smaller than the conventionally suggested $\mu_r=m_t$, reflecting the small virtuality of the QCD dynamics of the top-quark decay process. By including the NLO electroweak corrections, the finite $W$ boson width and the finite bottom quark mass, we obtain a reliable prediction for the top-quark total decay width
\begin{eqnarray}
\Gamma^{\rm tot}_t=1.3112^{+0.0188}_{-0.0187}\pm 0.0016\pm 0.0023 ~\rm {GeV},
\end{eqnarray}
where the errors are caused by the top-quark mass $m_t=172.5\pm0.7$ GeV, the coupling constant $\alpha_s(M_Z)=0.1179\pm0.0009$~\cite{PDG:2020} and the estimation of unknown higher-order terms using the PAA method with [N/M]=[1/1], respectively. The PMC improved predictions for the top-quark decay are helpful for detailed studies of properties of the top-quark. The current PMC results are complementary to the previous PMC calculations for top-quark pair production.

\hspace{1cm}

{\bf Acknowledgements}: This work was supported in part by the Natural Science Foundation of China under Grants No.12265011, No.11705033, No.11905056, No.12175025 and No.12147102; by the Project of Guizhou Provincial Department under Grant No.KY[2021]003 and [2020]1Y027.

\end{document}